\documentclass{mn2e}
\usepackage{epsfig,natbib2,natbibmnfix}
\usepackage{amssymb}
\usepackage{amssymb}
\usepackage{multirow}
\usepackage{multicol}
\usepackage[varg]{txfonts}
\usepackage{color}
\usepackage[totalwidth=480pt, totalheight=680pt]{geometry}

\newcommand{\cref}[1]{Chapter~\ref{#1}}

\begin{document}
\title[The distribution of metals in hot DA white dwarfs]{The distribution of metals in hot DA white dwarfs}

\author[N.J.~Dickinson,  M.A.~Barstow \& I. ~Hubeny] 
{
\parbox{5in}{N.J. Dickinson$^{1}$, M.A. Barstow$^{1}$ and I. Hubeny$^{2}$}
\vspace{0.1in} 
 \\ $^{1}$Department of Physics \& Astronomy, University of Leicester,
 Leicester LE1 7RH. UK.
\\
 $^{2}$Steward Observatory, The University of Arizona, 933 N. Cherry Ave, Tucson, AZ 85721. USA.
}

\maketitle

\begin{abstract}
The importance to stellar evolution of understanding the metal abundances in hot white dwarfs is well known. Previous  work has found the hot DA white dwarfs REJ 1032+532, REJ 1614-085 and GD 659 to have highly abundant, stratified photospheric nitrogen, due to the narrow absorption line profiles of the FUV N V doublet and the lack of EUV continuum absorption. A preliminary analysis of the extremely narrow, deep line profiles of the photospheric metal absorption features of PG 0948+534 suggested a similar photospheric metal configuration. However, other studies have found REJ 1032+532, REJ 1614-085 and GD 659 can be well described by homogeneous models, with nitrogen abundances more in keeping with those of white dwarfs with higher effective temperatures. Here, a re-analysis of the nitrogen absorption features seen in REJ 1032+532, REJ 1614-085 and GD 659 is presented, with the aim of better understanding the structure of these stars, to test which models better represent the observed data and apply the results to the line profiles seen in PG 0948+534. A degeneracy is seen in the modelling of the nitrogen absorption line profiles of REJ 1032+532, REJ 1614-085 and GD 659, with low abundance, homogeneously distributed nitrogen models most likely being a better representation of the observed data. In PG 0948+534, no such degeneracy is seen, and the enigmatically deep line profiles could not be modelled satisfactorially.

\end{abstract}


\footnotetext[1]{E-mail: njd15@le.ac.uk}

\section{Introduction}
\label{intro}

As the end products of the lives of the vast majority of stars, a thorough understanding of white dwarfs is crucial to our understanding of stellar evolution. To understand the evolution of white dwarfs, accurate measurements of parameters such as the effective temperature of a star (\textit{T}$_{\rm eff}$, which gives how far along the cooling sequence the white dwarf has travelled) are required. A common technique (pioneered by \citealt{Holbergetal85} and extended to a large white dwarf sample by \citealt{BergeronSafferLiebert92}) used to measure a white dwarf \textit{T}$_{\rm eff}$ is to compare the observed Balmer line profiles to model calculations. The surface gravity (log \textit{g}) of the star can be found in a similar way. Using these parameters as inputs to white dwarf evolutionary models (such as those of \citealt{Wood95}), one can then derive a stellar mass, which can then be used in studies of white dwarf mass distributions (e.g. \citealt{BergeronSafferLiebert92,LiebertBergeronHolberg05}), luminosity functions (e.g. \citealt{LiebertBergeronHolberg05}) or initial-final mass relations (e.g. \citealt{Casewelletal09,Dobbieetal09}). Being among the oldest stellar objects, white dwarfs can be used as chronometers to age stellar populations (e.g. \citealt{FontaineBrassardBergeron01}); the age of the oldest white dwarf in the galactic disk can provide a lower age estimate for the Milky Way. Reliable measurements of \textit{T}$_{\rm eff}$ are therefore critical to our understanding of white dwarf stars and their application to stellar evolution.

For some time, it has been known that metals are present in the photospheres of almost all hot DA white dwarfs (e.g. \citealt{Barstowetal93,Marshetal97}) due to radiative levitation (e.g. \citealt{Chayeretal94,Chayeretal95}). As first suggested by \cite{DreizlerWerner93}, the line blanketing caused by photospheric metals significantly affects \textit{T}$_{\rm eff}$ measurements for DA white dwarfs with \textit{T}$_{\rm eff}$ $>$ 55000 K (\citealt{BarstowHubenyHolberg98}). Also, \cite{Barstowetal01,Barstowetal03BalLy} found that for \textit{T}$_{\rm eff}$ $>$ 50000 K (the \textit{T}$_{\rm eff}$ at which radiative levitation effects dominate and are responsible for putting metals into DA white dwarf photospheres) a significant difference between \textit{T}$_{\rm eff}$ values derived from Balmer and Lyman line measurements arises (a similar effect is seen in DAO stars, \citealt{Goodetal04}). Uncertainties in modelling the metals in hot white dwarf photospheres (an analysis of the metal abundance patterns in hot DA white dwarfs shows that although the abundance patterns predicted by radiative levitation are broadly reproduced, the precise, measured abundances do not often match those predicted, \citealt{Barstowetal03patt}) could go some way to explaining this phenomenon. Clearly, in light of the importance of accurate white dwarf \textit{T}$_{\rm eff}$ measurements, a rigorous understanding of the metals in hot white dwarfs is required. 

\begin{figure}
  \includegraphics[width=0.47\textwidth]{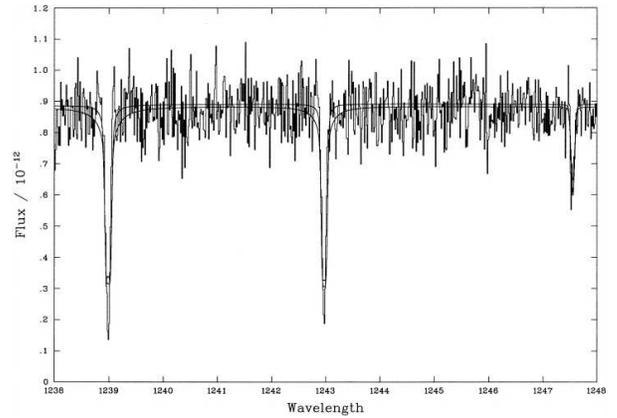}
  \caption{Figure 5 from  \cite{Holbergetal99ph}. The N V doublet of REJ 1032+532 fit with a homogeneous model with N(\textit{N})/N(\textit{H})=5x10$^{-5}$ is represented by the lower curve. The upper curve is a model with stratified ($\Delta$M/M=3.1x10$^{-16}$) nitrogen (again, N(\textit{N})/N(\textit{H})=5x10$^{-5}$). A C III absorption feature is also seen near 1247.5 \AA.}
  \label{fig:rej1032FUVh99}
\end{figure}

\begin{figure}
  \includegraphics[width=0.47\textwidth]{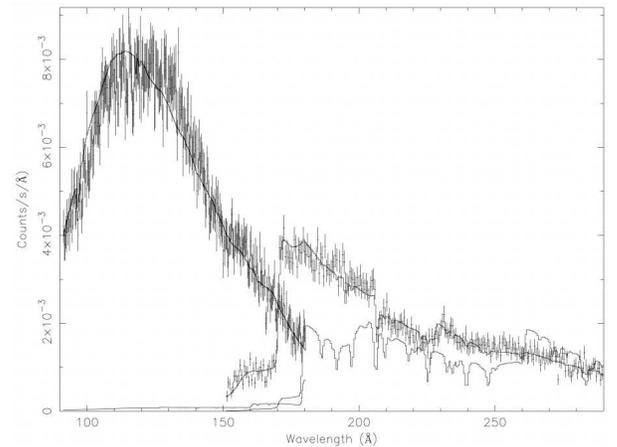}
  \caption{Figure 6 from \cite{Holbergetal99ph}. The data are the EUVE SW and MW EUV continuum of REJ 1032+532. The best fitting models for each dataset are the stratifed nitrogen model. The other poorly fitting, heavily absorbed models are those with homogeneous nitrogen (again, a model is pictured for both the SW and MW data).}
  \label{fig:rej1032EUVh99}
\end{figure}

\begin{figure}
  \includegraphics[width=0.47\textwidth]{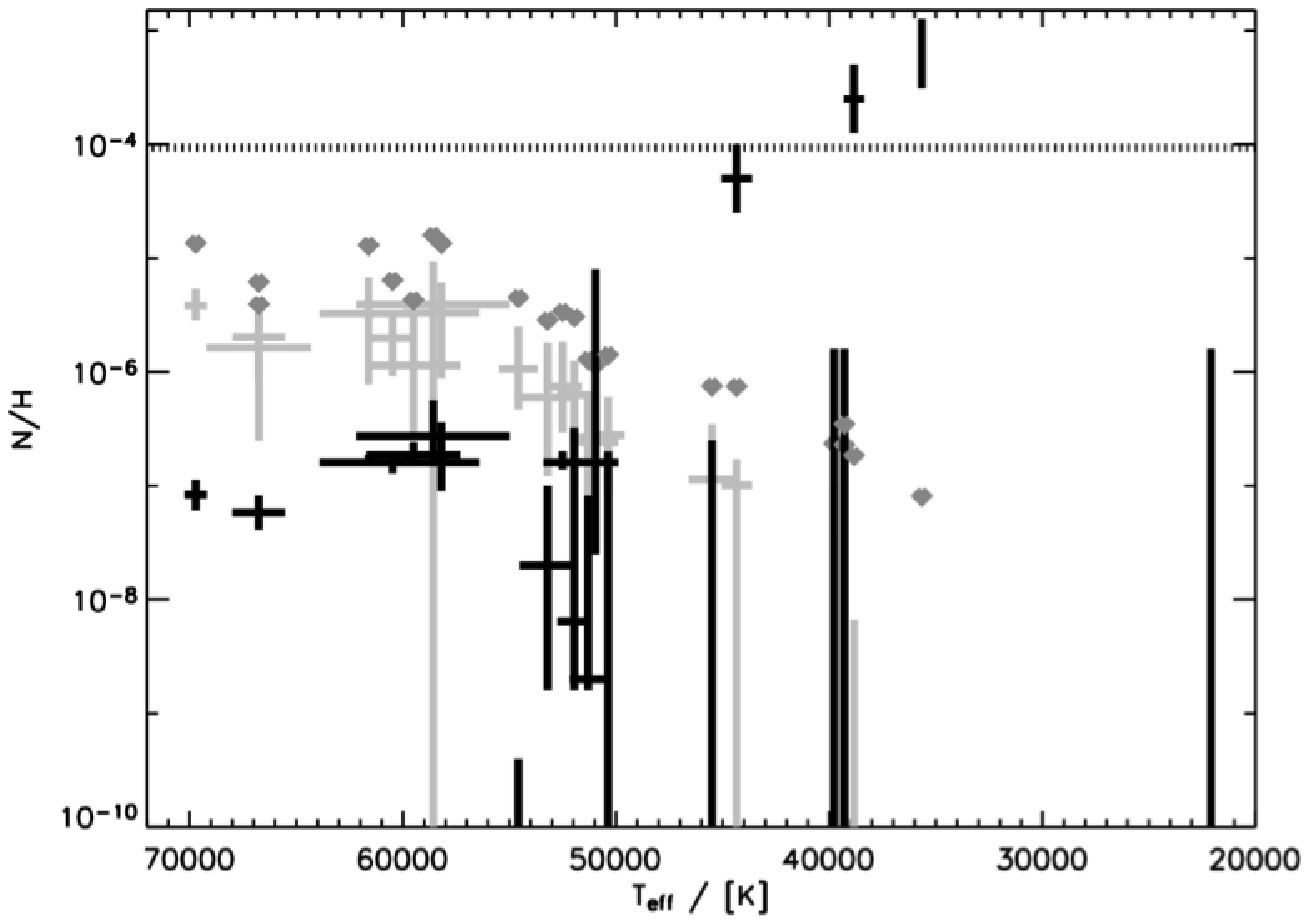}
  \caption{Figure 1 (third row, left hand panel) from \cite{SchuhBarstowDreizler05}. A comparison of the nitrogen abundances measured by \cite{Barstowetal03patt} (black symbols) to the measurements of \cite{SchuhBarstowDreizler05} (light grey symbols). The dark grey symbols show the radiative levitation predictions of \cite{Chayeretal95} and the dotted line shows the cosmic abundance. Note that when non-detections of nitrogen were made by \cite{Barstowetal03patt}, an upper limit was estimated, leading to the large error bars seen in the black data points at lower \textit{T}$_{\rm eff}$.}
  \label{fig:schuhbarstow}
\end{figure}

As well as the stratified hydrogen/helium configurations well known in many DAs (e.g. \citealt{Vennesetal88}), evidence for stratified metal distributions (a phenomenon predicted by \citealt{Chayeretal95}) has been seen in some stars. In a study of G191-B2B, \cite{BarstowHubenyHolberg98} found a inhomogeneous iron distribution best fit the EUV, FUV and optical observations of the star, though the predicted and observed fluxes disagreed in the shortest wavelength regime ($\lambda$ $<$ 190 \AA). The stratified iron configuration was found to have an increasing abundance with depth; the iron depletion in the upper atmosphere was suggested to be a consequence of radiatively driven mass loss. \cite{Dreizler99} developed a method of modelling the EUV spectral region of hot white dwarfs, where the chemical abundance at each depth point was that produced by the equilibrium between radiative levitation and downward diffusion (with depth dependent radiation intensity). This gave a depth dependent abundance distribution, and successfully modelled the EUV spectrum of G191-B2B (\citealt{DreizlerWolff99}). A stratified nitrogen distribution was also used to simultaneously explain the FUV N V doublet (1238.82, 1242.80 \AA\AA) and the observed EUV continuum in REJ 1032+532 (\citealt{Holbergetal99ph}). Homogeneously distributed nitrogen with an abundance of N(\textit{N})/N(\textit{H})=5x10$^{-5}$ gave model line profiles with a depth similar to those observed, though they were pressure broadened beyond the observed profiles (figure \ref{fig:rej1032FUVh99}, lower curve). Confining the nitrogen to a layer at the top of the atmosphere ($\Delta$M/M=3.1x10$^{-16}$) reduced this pressure broadening while maintaining the depth of the observed line profiles, better matching the data (figure \ref{fig:rej1032FUVh99}, upper curve). The homogeneous nitrogen model caused heavy absorption of the EUV continuum (figure \ref{fig:rej1032EUVh99}, lower curve); the lack of nitrogen in the lower atmosphere in the stratified model removed this absorber and reproduced the observed EUV continuum (figure \ref{fig:rej1032EUVh99}, upper curve). Using \textit{FUSE} observations of the O VI doublet (1031.912, 1037.613 \AA\AA) and \textit{EUVE} data, \cite{Chayeretal06} found REJ 1032+532 to have a similarly stratified oxygen distribution (no interstellar O VI was observed by \citealt{Barstowetal10} to contaminate the photospheric O VI line profiles). Stratified nitrogen distributions similar to that of REJ 1032+532 were suggested to explain the N V doublet and EUV observations of GD 659 (\citealt{Holbergetal95,Barstowetal03patt}) and the N V doublet of REJ 1614-085 (\citealt{Holbergetal97,Holbergetal00,Barstowetal03patt}). The analysis of \cite{Barstowetal03patt} found that above 50000 K there was no pattern between nitrogen abundance and \textit{T}$_{\rm eff}$. Below 50000 K a dichotomy was observed, with the three white dwarfs discussed here forming a trend of increasing nitrogen abundance with decreasing \textit{T}$_{\rm eff}$, while definite nitrogen detections in the other white dwarfs could not be made (figure \ref{fig:schuhbarstow}, black symbols).

\begin{figure}
  \includegraphics[width=0.47\textwidth]{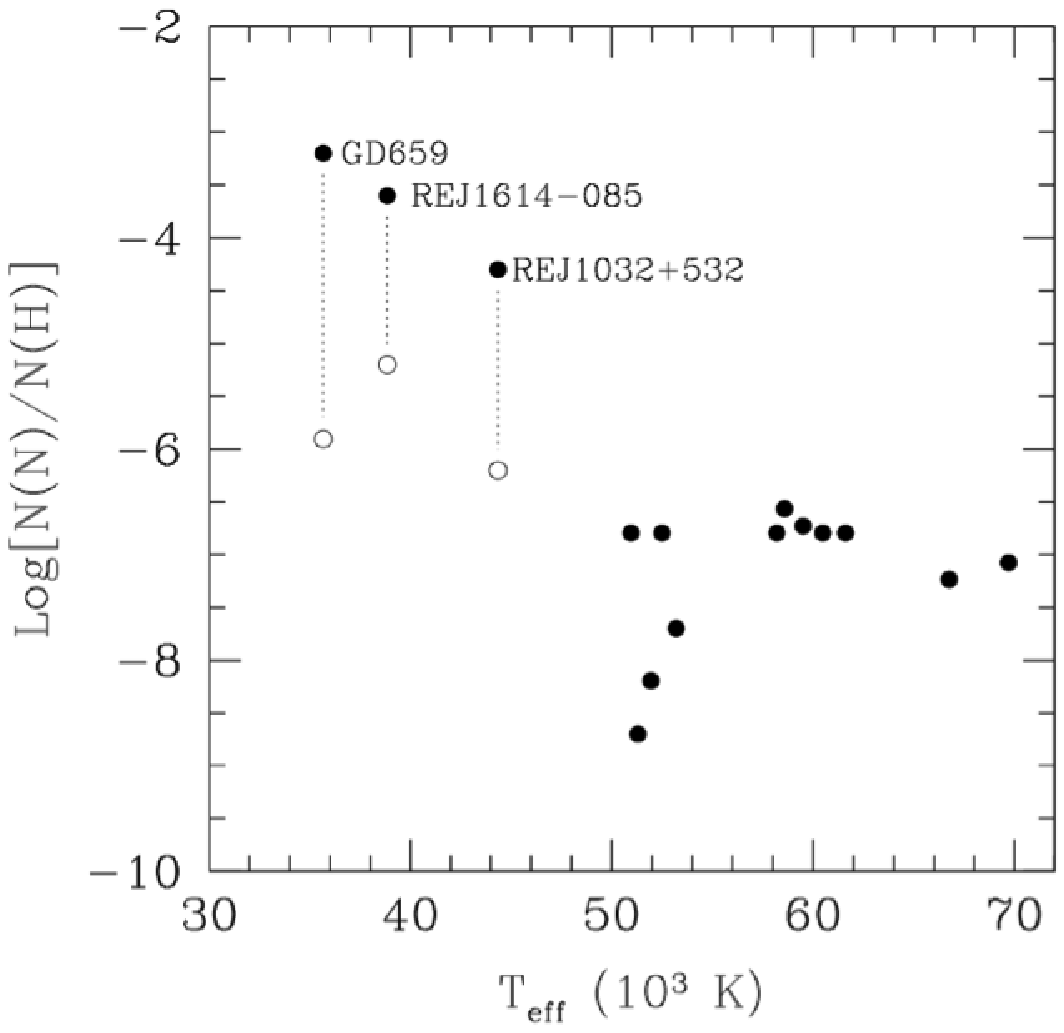}
  \caption{Figure 1 (left hand panel) from \cite{Chayeretal05}. A comparison of the nitrogen abundances measured by \cite{Barstowetal03patt} (filled circles) to the measurements of \cite{Chayeretal05} (open circles).}
  \label{fig:chayerbarstow}
\end{figure}

\cite{SchuhDreizlerWolff02} further developed and applied the stratified modelling method of \cite{Dreizler99} and \cite{DreizlerWolff99} to obtain metal abundances for a sample of stars. All stars were well fit with stratified metal distributions, save a few exceptions. Among these exceptions was REJ 1032+535 (= REJ 1032+532), which was better fit with a homogeneous metal distribution. A comparison by \cite{SchuhBarstowDreizler05} of the abundances measured using the stratified modelling method of \cite{SchuhDreizlerWolff02}  (figure \ref{fig:schuhbarstow}, light grey symbols) to the measurements obtained from the homogeneous model grids of \cite{Barstowetal03patt} (figure \ref{fig:schuhbarstow}, black symbols) found that the nitrogen abundances for REJ 1032+532, REJ 1614-085 and GD 659 were much lower than those obtained by \cite{Barstowetal03patt}. The trend of increasing nitrogen abundance with decreasing \textit{T}$_{\rm eff}$ observed by \cite{Barstowetal03patt} was also not observed. \cite{Chayeretal05} came to a similar conclusion (figure \ref{fig:chayerbarstow}), though the measured abundances in this case were noticeably greater than those found by \cite{SchuhBarstowDreizler05} in the cases of REJ 1614-085 and GD 659.  \cite{Chayeretal05} used both FUV and EUV data and homogeneous model grids calculated using the same method used by \cite{Barstowetal03patt}. Indeed, a remark made by \cite{Schuh05} was that it was of concern that both \cite{Barstowetal03patt} and \cite{Chayeretal05} obtained such different results using the same method.

Given the conflicting results outlined so far, a further analysis to better constrain the distribution and abundance of nitrogen in REJ 1032+532, REJ 1614-085 and GD 659 is desirable, and is presented here. The absorption line profiles of PG 0984+534 are extremely narrow and deep, akin to those of the REJ 1032+532 N V doublet. \cite{Barstowetal03patt} stated that preliminary calculations suggested that the metals in the photosphere of the star was also stratified and, in light of the investigation into the photospheric nitrogen in REJ 1032+532, REJ 1614-085 and GD 659, PG 0948+534 is also investigated to see whether its enigmatic absorption line profiles can be modelled using stratified metal configurations.

\section{Observations and Method}
\label{method}

The observations used in this work are detailed in table \ref{table:tloggmet}. Though the method applied here has been described in detail in previous work (e.g. \citealt{Barstowetal03}), a summary is presented here. Models were computed using  the non-LTE code TLUSTY (e.g. \citealt{HubenyLanz95}). The \textit{T}$_{\rm eff}$ and log \textit{g} values measured by \cite{Barstowetal03} were used in the model calculations, and can be found in table \ref{table:tloggmet} (along with the metal abundances measured by \citealt{Barstowetal03}).  The C, O, and Si abundances found by \cite{Barstowetal03} were used in the model calculations of REJ 1032+532, REJ 1614-085 and GD 659 (where both C III and C IV abundances were found by \cite{Barstowetal03}, the higher abundance was included). Model atoms for C III, C IV, N III, N IV, N V, O IV, O V, O VI, Si III and Si IV were explicitly included, while C V, N VI, O VII and Si V were treated as one level atoms. The hydrogen broadening tables of \cite{Lemke97} were also included. Unlike the study of \cite{Barstowetal03}, which probed only the high nitrogen abundance regime, homogeneous model grids were constructed for each object to encompass the nitrogen abundances of both \cite{Barstowetal03} and \cite{Chayeretal05}, to allow a full examination of the variation of the N V doublet line profile and EUV continuum with nitrogen abundance.

\begin{table*}
\caption{The observation information, \textit{T}$_{\rm eff}$, log \textit{g}, metal abundances and ISM column densities towards the white dwarfs studied here. All observations, values of \textit{T}$_{\rm eff}$, log \textit{g} and metal abundances are those detailed by \cite{Barstowetal03}, unless stated otherwise. The interstellar column densities are from the references indicated. The absence of data signifies where a measurement was unobtainable, either due to lack of spectral coverage in the required region or an inability to model absorption features.}
\begin{tabular}{l c c c c}
\hline
                               & PG 0948+534                    & REJ 1032+532                          & REJ 1614-085                     & GD 659\\
\hline
FUV data                       & HST \textit{STIS}[E140M]       & HST \textit{STIS} [E140M]$^{a}$       & HST \textit{GHRS} [G160M]$^{b}$  & HST \textit{STIS} [E140M]\\
EUV data                       &                                & \textit{EUVE} [SW]$^{a}$       &                                  & \textit{EUVE} [SW]$^{c}$\\
\textit{T}$_{\rm eff}$ (K)      & 110000                       & 44350                               & 38840                          & 35660\\
log \textit{g}                  & 7.58                           & 7.81                                  & 7.92                             & 7.93 \\ 
\textit{N}(C III)/\textit{N}(H) &                                & 3.00x10$^{-7}$                        &                                  & 0.00\\
\textit{N}(C IV)/\textit{N}(H)  &                                & 1.60x10$^{-7}$                        & 7.00x10$^{-7}$                   & 5.00x10$^{-8}$\\
\textit{N}(N V)/\textit{N}(H)   &                            & 5.00x10$^{-5}$                        & 2.50x10$^{-4}$                   & 6.30x10$^{-4}$\\
\textit{N}(N V)/\textit{N}(H)$^{d}$ &                            & 1.26x10$^{-6}$                        & 6.31x10$^{-7}$                   & 6.31x10$^{-6}$\\
\textit{N}(O V)/\textit{N}(H)   &                                & 1.20x10$^{-7}$                        &                                  & 0.00\\
\textit{N}(Si IV)/\textit{N}(H) &                                & 9.50x10$^{-7}$                        & 9.50x10$^{-9}$                   & 4.80x10$^{-9}$\\
\textit{N}(Fe V)/\textit{N}(H)  & 1.90x10$^{-6}$                 & 0.00                                  & 0.00                             & 0.00\\
\textit{N}(Ni V)/\textit{N}(H)  & 1.20x10$^{-7}$                 & 0.00                                  & 0.00                             & 0.00\\
log(\textit{N}$_{\rm H I}$) (cm$^{-2}$)         &                                & 18.62$^{e}$                           &                                  & 18.46$^{f}$\\
log(\textit{N}$_{\rm He I}$) (cm$^{-2}$)       &                                & 17.75$^{e}$                           &                                  & 17.37$^{f}$ \\
log(\textit{N}$_{\rm He II}$) (cm$^{-2}$)       &                                & 17.28$^{e}$                           &                                  & 17.17$^{f}$ \\

\hline
\end{tabular}
\\$^{a}$\cite{Holbergetal99ph}, $^{b}$\cite{Holbergetal97},$^{c}$\cite{Holbergetal95}, $^{d}$\cite{Chayeretal05}, $^{e}$\cite{Holbergetal99is}, $^{f}$\cite{Barstowetal97}.
\label{table:tloggmet}
\end{table*}

FUV spectra were synthesised from 1235\AA\ to 1245\AA\ to cover the N V doublet, and the XSPEC package was used to compare the model spectra to the observed data using a $\chi^{2}$ minimisation technique. When a best fit nitrogen abundance was found, a set of stratifed models were calculated, with the nitrogen abundance at the top of the atmosphere equal to the best fitting homgeneous abundance. Below this layer the nitrogen abundance was zero. In each model, the depth of this nitrogen layer was extended, allowing a model grid to be constructed to investigate whether a stratified metal configuration better matched the data (it must be stressed that the stratified models calculated here are not the self-consistently formulated, depth dependent models of the form of \citealt{Dreizler99}, \citealt{DreizlerWolff99} and \citealt{SchuhDreizlerWolff02}, but are `exploratory' stratified models of the type computed by \citealt{BarstowHubenyHolberg98} and \citealt{Holbergetal99ph}). The EUV analysis was conducted in the SW range (80 \AA\ to 180 \AA), since the EUV absorption was most severe there. Again, XSPEC was used to compare the models to the data. EUV interstellar absorption along the sight lines to the white dwarfs was accounted for using the column densities in table \ref{table:tloggmet}. The model grids used by \cite{Barstowetal03} were used to fit the FUV absorption features (table \ref{table:linelist}) of PG 0948+534. After initial metal abundances were estimated, stratified grids were computed in the same way as the stratified nitrogen grids. The abundances of the metals other than the metal being stratified were fixed at the abundance estimates obtained using the homogeneous models computed by \cite{Barstowetal03}. The Fe and Ni abundances measured by \cite{Barstowetal03} were included in all models of this object.

\begin{table}
\caption{The high ions examined in the spectrum of PG 0948+534 and their laboratory wavelengths}
\begin{tabular}{c c}
\hline
Ion & Lab. Wavelength (\AA) \\
\hline
C IV   & 1548.187, 1550.772 \\
N V    & 1238.821, 1242.804 \\
O V    & 1371.296 \\
Si IV  & 1393.755, 1402.770 \\
\hline
\end{tabular}
\label{table:linelist}
\end{table}
\section{REJ 1032+532}
\label{REJ 1032+532}

\begin{figure}
  \includegraphics[width=0.47\textwidth]{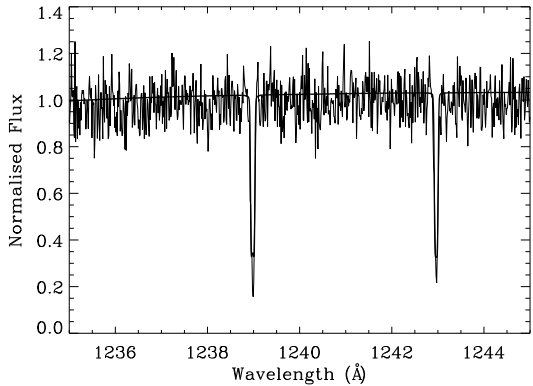}
  \caption{The FUV N V doublet of REJ 1032+532, fit with a homogeneous model with \textit{N}(N)/\textit{N}(H) = 3.39x10$^{-7}$.}
  \label{rej1032FUV}
\end{figure}

\begin{figure}
\includegraphics[width=0.47\textwidth]{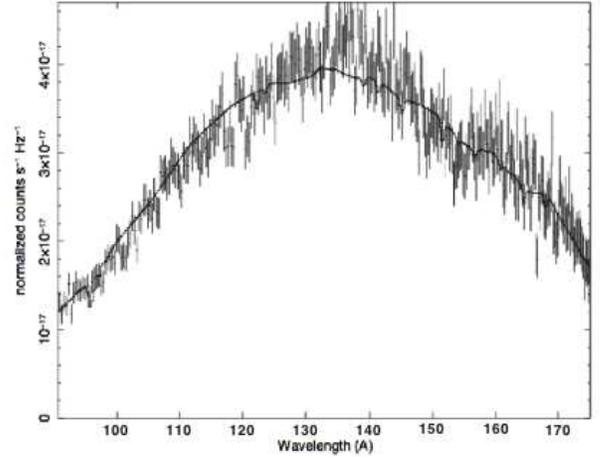}
\caption{The homogeneous model of REJ 1032+532, with \textit{N}(N)/\textit{N}(H) = 3.39x10$^{-7}$, fit to the \textit{EUVE} SW data.}
  \label{fig:1032EUV}
\end{figure}

An examination of the FUV N V doublet (1238.821, 1242.804 \AA\AA) found a global best fit for the homogeneous model at 3.39$\pm^{1.29}_{1.31}$x10$^{-7}$, with a $\chi$$_{\nu}^{2}$ = 0.58 (figure \ref{rej1032FUV}). A secondary best fit was found at 5.00$\pm^{2.55}_{0.48}$x10$^{-5}$, with a $\chi$$_{\nu}^{2}$ = 0.77. Stratifying the model at either abundance did not improve the fit over the lower abundance, homogeneous model. The lower abundance model fit the EUV data well (figure \ref{fig:1032EUV}); no nitrogen absorption edge was seen in the SW data, where it had dominated in the high abundance homogeneous model of \cite{Holbergetal99ph} (figure \ref{fig:rej1032EUVh99}).

\section{REJ 1614-085}
\label{REJ 1614-085}

\begin{figure}
  \includegraphics[width=0.47\textwidth]{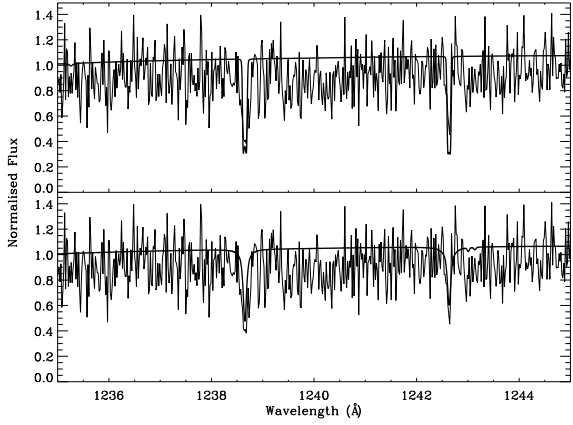}
  \caption{The lower abundance model (1.76x10$^{-6}$; $\chi$$_{\nu}^{2}$ = 1.21) of REJ 1614-085 is in the upper panel. The high nitrogen abundance model (3.41x10$^{-4}$; $\chi$$_{\nu}^{2}$ = 1.13) is shown in the lower panel. }
 \label{fig:rej1614FUV}
\end{figure}

A lower nitrogen abundance secondary best fit was found at 1.76$\pm^{1.65}_{1.26}$x10$^{-6}$ ($\chi$$_{\nu}^{2}$ = 1.21) (figure \ref{fig:rej1614FUV}, upper panel). Unlike REJ 1032+532, the analysis of the FUV N V doublet of this DA found a best fitting abundance in the high abundance regime, at 3.41$\pm^{1.81}_{1.50}$x10$^{-4}$ ($\chi$$_{\nu}^{2}$ = 1.13) (figure \ref{fig:rej1614FUV}, lower panel). The lower abundance model here over predicted the depths of the observed line profiles, while the absorption lines of the higher abundance model did not quite extend into the observed line profiles. An explanation of this behaviour and the degeneracy in the previous star is presented in section \ref{discussion}.

\section{GD 659}
\label{GD 659}

\begin{figure}
  \includegraphics[width=0.47\textwidth]{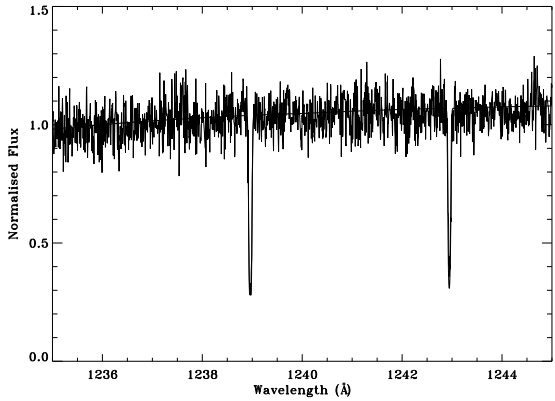}
  \caption{The FUV N V doublet of GD 659, fit with a homogeneous model with \textit{N}(N)/\textit{N}(H) = 6.05x10$^{-7}$.}
  \label{fig:gd659FUV}
\end{figure}

In keeping with REJ 1032+532, two best fitting abundances were found for GD 659, with the overall best fitting abundance at 6.05$\pm^{0.64}_{0.62}$x10$^{-7}$  ($\chi$$_{\nu}^{2}$ = 2.15; figure \ref{fig:gd659FUV}) and the secondary best fitting abundance at 5.70$\pm^{0.41}_{0.62}$x10$^{-5}$ ($\chi$$_{\nu}^{2}$ = 2.41). Again, stratifying the high abundance model did not offer a better fit to the data than the lower abundance, homogeneous model. Like REJ 1032+532, no nitrogen absorption edge was seen at EUV wavelengths at the lower, homogeneous abundance.

\section{PG 0948+534}
\label{PG 0948+534}
\begin{figure}
  \includegraphics[width=0.47\textwidth]{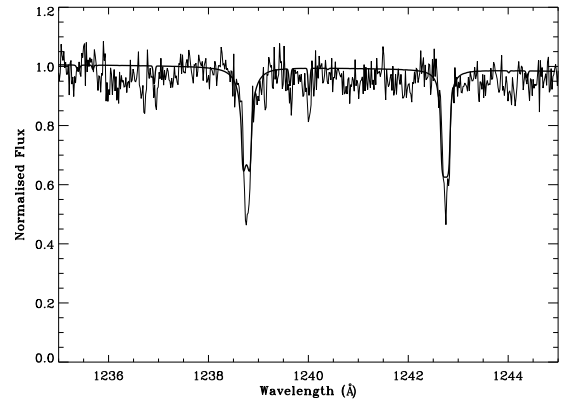}
  \caption{The N V doublet of PG 0948+534, fit with a model with a nitogen abundance of 1.60x10$^{-6}$.}
 \label{fig:pg0948}
\end{figure}

The absorption line profiles in the spectrum of PG 0948+534 could not be modelled satisfactorially here (figure \ref{fig:pg0948}). The best fitting, homogeneous C IV abundance was found to be 4.85x10$^{-6}$, while the N V doublet, O V line  and Si IV doublet were best fit with abundances of 1.6x10$^{-6}$, 3.5x10$^{-5}$ (the O V grid upper limit of 100 times the abundance of G191-B2B) and 3.15x10$^{-5}$, respectively. Above these abundances, the depth of the line profiles could not be reproduced while maintaining the narrow width of the lines. Stratifying the metals did not improve the fits. Given the poor match to the data provided by these models, errors were not computed for these measurements.

\section{Discussion}
\label{discussion}

\begin{figure}
  \includegraphics[width=0.47\textwidth]{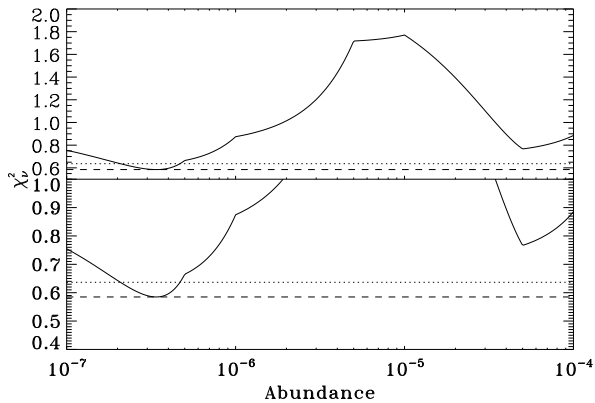}
  \caption{The $\chi$$_{\nu}^{2}$ distribution of REJ 1032+532 as nitrogen abundance increased. The global $\chi$$_{\nu}^{2}$ minimum is represented with a dashed line, and its 3$\sigma$ confidence limit is denoted with a dotted line.}
 \label{fig:rej1032chisq}
\end{figure}

A degeneracy was seen in the modelling of photospheric nitrogen in REJ 1032+532, REJ 1614-085 and GD 659. The best fitting abundances for each star are presented in table \ref{table:results} (the results for PG 0948+534 are not included as they are thought not to be reliable). Figure \ref{fig:rej1032chisq} shows the $\chi$$_{\nu}^{2}$ distribution of REJ 1032+532 as nitrogen abundance was increased. The upper panel shows that as \textit{N}(N)/\textit{N}(H) is increased, two minima are seen at the best fitting abundances of 3.39x10$^{-7}$ and 5x10$^{-5}$. The lower panel shows the minima in more detail, with the dashed line representing the global minimum in $\chi$$_{\nu}^{2}$ (0.58) and the dotted line showing the 3$\sigma$ confidence limit of this minimum. The secondary minimum ($\chi$$_{\nu}^{2}$ = 0.77) is clearly separated from the confidence limit of the global minimum, so the global minimum can confidently be declared the best fitting abundance. The lower abundance model also fit the \textit{EUVE} SW data well. A similar situation occurrs for GD 659, with  a best fit abundance of 6.05$\pm^{0.64}_{0.62}$x10$^{-7}$.

\begin{table}
\caption{The nitrogen abundances for REJ 1032+532, REJ 1614-085 and GD 659.}
\begin{tabular}{l c c c}
\hline
Star         & \textit{N}(N)/\textit{N}(H) & +3$\sigma$        & -3$\sigma$\\
\hline
REJ 1032+532 & 3.39x10$^{-7}$              & 1.29x10$^{-7}$  & 1.31x10$^{-7}$\\
REJ 1614-085 & 1.76x10$^{-6}$              & 1.65x10$^{-7}$  & 1.26x10$^{-7}$\\
GD659        & 6.05x10$^{-7}$              & 0.64x10$^{-7}$  & 0.62x10$^{-7}$\\
\hline
\end{tabular}
\label{table:results}
\end{table}

\begin{figure}
  \includegraphics[width=0.47\textwidth]{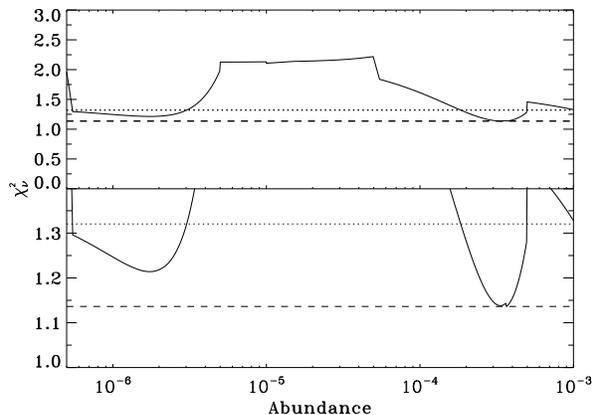}
  \caption{The $\chi$$_{\nu}^{2}$ distribution of REJ 1614-085. Again, the global $\chi$$_{\nu}^{2}$ minimum is represented with a dashed line, and its 3$\sigma$ confidence limit is denoted with a dotted line.}
 \label{fig:rej1614chisq}
\end{figure}

In REJ 1614-085, the two minima are less easy to disentangle. The lower panel of figure \ref{fig:rej1614chisq} shows that the separation of the two minima is not greater than the 3$\sigma$ confidence limit of the global minimum. Indeed, the higher nitrogen abundance (3.41$\pm^{1.81}_{1.50}$x10$^{-4}$) is, marginally, the global minimum. A close inspection of the $\chi$$_{\nu}^{2}$ distribution shows that poorly defined minima are also present near  \textit{N}(N)/\textit{N}(H) = 10$^{-5}$ (figure \ref{fig:rej1614chisq}, upper panel) and 7x10$^{-3}$ (figure \ref{fig:rej1614chisq}, lower panel). Given that both REJ 1032+532 and GD 659 are well explained with lower nitrogen abundance, homogeneous models and that the minima in the $\chi$$_{\nu}^{2}$ distribution of REJ 1614-085 cannot be confidently disentangled, the lower nitrogen abundance (1.76$\pm^{1.65}_{1.26}$x10$^{-6}$) is deemed the best fit here. Better signal-to-noise data may help to break this degeneracy with more confidence, and may allow a better match to the observed line profiles. In PG 0948+534, the  $\chi$$_{\nu}^{2}$ distributions for C IV, N V, O V and Si IV have single minima. The values of $\chi$$_{\nu}^{2}$ were high (with values of 2.9-9.8 across the high ion fits), showing that the models used here do not appropriately represent the metal absorption seen in this star. 

\begin{figure}
  \includegraphics[width=0.47\textwidth]{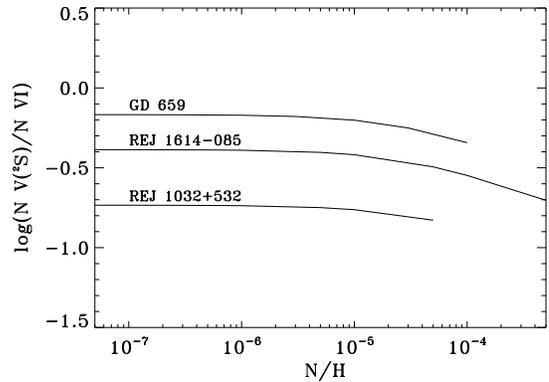}
  \caption{The change in the log of the ratio of the N V NLTE populations responsible for the N V doublet to the N VI level population in the stellar models, with nitrogen abundance.}
 \label{fig:nvnvi}
\end{figure}

A comparison of the model NLTE population responsible for the N V doublet absorption (${^{2}}$S) to the N VI level shows that at high nitrogen abundances, the value of log(N V(${^{2}}$S)/N VI) decreases above nitrogen abundances $\sim$10$^{-6}$ (figure \ref{fig:nvnvi}. Note that not all the model grids cover the same abundance range, only the range required to explain the observed line profiles in each case.). This `over-ionisation' may provide a mechanism for the apparent degeneracy; although the nitrogen abundance in each cases is increasing, the relative amount of N V (${^{2}}$S) decreases. Given that log(N V(${^{2}}$S)/N VI) of REJ 1614-085 follows the same trend as the log(N V(${^{2}}$S)/N VI) of both REJ 1032+532 and GD 659, this further suggests that though the degeneracy seen in figure \ref{fig:rej1614chisq} cannot be confidently broken, REJ 1614-085 has a homogeneous nitrogen distribution with an abundance in keeping with the other objects.

The work (e.g. \citealt{Holbergetal99ph}) that found the nitrogen in these stars to be stratified and of high abundance cite a radiatively driven mass loss process as the reason for the nitrogen distribution. Recent studies looking at mass loss in DA white dwarfs have found that for DAs in the 25000 K $<$ \textit{T}$_{\rm eff} <$ 50000 K range, mass loss cannot occur for stars with log \textit{g} $>$ 7.0 since the upward radiative pressure cannot overcome the gravity of the star (\citealt{Unglaub08}). Indeed, even up to and beyond \textit{T}$_{\rm eff}$ = 60000 K, mass loss cannot occur for DAs with log \textit{g} $>$ 7.0 (\citealt{UnglaubBues00,Unglaub07}). No circumstellar N V has been found at these objects, from which accretion can take place and enrich the upper atmosphere (\citealt{Bannisteretal03,Dickinsonetal12}). This suggests that the nitrogen enrichment in the upper atmosphere due to either mass loss or accretion cannot be happening, strengthening the argument for a homogeneous nitrogen distribution in all of the objects. 

\begin{figure}
  \includegraphics[width=0.47\textwidth]{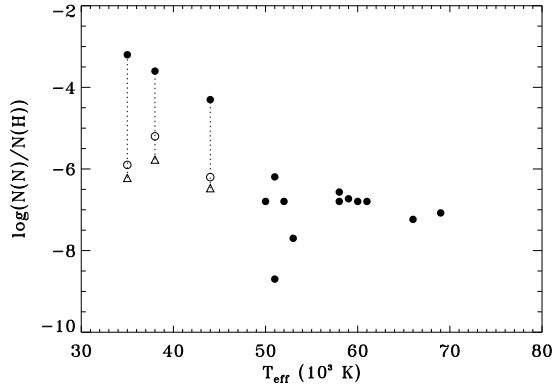}
  \caption{A comparison of the nitrogen abundances found here (triangles) to those found by \cite{Barstowetal03} (filled circles) and \cite{Chayeretal05} (open circles). The dotted lines connect the data points of a given star to aid comparison.}
 \label{fig:comparebarstow}
\end{figure}

\begin{figure}
  \includegraphics[width=0.47\textwidth]{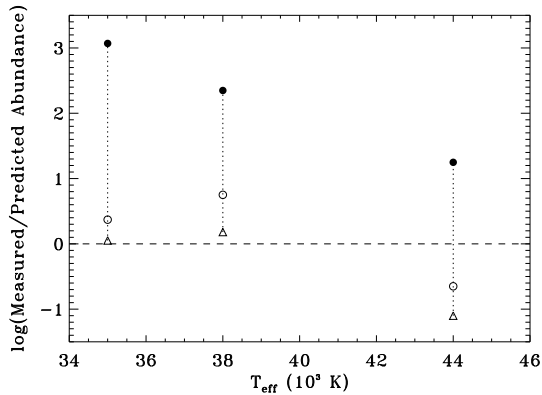}
  \caption{A comparison of the abundances found here,by \cite{Barstowetal03} and  by \cite{Chayeretal05} to those predicted by \cite{Chayeretal95}. The plot symbols are the same as for figure \ref{fig:comparebarstow}}
 \label{fig:comparechayer95}
\end{figure}

Figure \ref{fig:comparebarstow} shows a comparison of the nitrogen abundances found here (triangular symbols) to those found by \cite{Barstowetal03} (filled circles) and \cite{Chayeretal05} (open circles). The nitrogen abundances measured here are more in keeping with the nitrogen abundances measured by \cite{Barstowetal03} for DA white dwarfs of higher \textit{T}$_{\rm eff}$. The pattern between nitrogen abundance and \textit{T}$_{\rm eff}$ seen by \cite{Chayeretal05} is reproduced, though the measured abundances here are offset to those measured by \cite{Chayeretal05} (probably due to systematic differences in the fitting procedures used by \citealt{Chayeretal05} when compared to the method here). The trend of increasing nitrogen abundance seen by \cite{Barstowetal03} is not seen here. A comparison of the results for REJ 1032+532, REJ 1614-085 and GD 659 to the radiative levitation predictions of \cite{Chayeretal95} (figure \ref{fig:comparechayer95}) shows that the abundances derived here are much closer to those those predicted by radiative levitation theory than those found by \cite{Barstowetal03}.

In view of the fact that the low nitrogen abundance models were preferred over the high nitrogen abundance models in the cases of REJ 1032+532, REJ 1614-085 and GD 659, it was perhaps no surprise that stratified model calculations also failed to explain the high ion absorption features of PG 0948+534. The much higher \textit{T}$_{\rm eff}$ of PG 0948+534 (being 65650 K hotter than the next hottest star, REJ 1032+532) may introduce some physical effects the are not accounted for in the models. Another possible reason for the observed line profiles in this object is the presence of unresolved material along the sight line of the white dwarf at a velocity co-incident with the photosphere. This will be explored in a later paper.

\section*{Acknowledgements}
\label{acknowledgements}
NJD and MAB acknowledge the support of STFC. NJD wishes to thank Klaus Unglaub and Matt Burleigh for useful discussions.
\bibliographystyle{mn2e} 
\bibliography{struct}

\end{document}